# Nutrients and biomass dynamics of photo-sequencing batch reactors treating wastewater with high nutrients loadings

Dulce María Arias*, Enrica Uggetti, María Jesús García-Galán, Joan García

[1]GEMMA – Group of Environmental Engineering and Microbiology, Department of Civil and Environmental Engineering, Universitat Politècnica de Catalunya•BarcelonaTech, c/ Jordi Girona 1-3, Building D1, E-08034, Barcelona, Spain.

* Corresponding author:

Tel.: +34 934015259

Fax: +34 934015259

E-mail address: dulce.maria.arias@upc.edu

**Abstract**

This study investigates different strategies as treatment of digestate from anaerobic digester diluted with the secondary effluent from a high rate algal pond. To this aim, the performance of two photo-sequencing batch reactors (PSBRs) operated at high nutrients loading rates and different solids retention times were compared with a semi-continuous photobioreactor (SC). Performances were evaluated in terms of wastewater treatment, biomass composition and polymers accumulation during 30 days of operation. PSBR$_{2-10}$ and PSBR$_{2-5}$ were operated at a hydraulic retention time (HRT) of 2 days with solids retention time (SRT) of 10 and 5, respectively, while semi-continuous reactor (SC$_{10-10}$) was operated at a coupled HRT/SRT of 10 days. Results showed that PSBR$_{2-5}$ achieved the highest removal rates in terms of TN (6.7 mg L$^{-1}$·d$^{-1}$), TP (0.31 mg L$^{-1}$·d$^{-1}$), TOC (29.32 mg L$^{-1}$·d$^{-1}$) and TIC (3.91 mg L$^{-1}$·d$^{-1}$). Those results were in general 3-6 times higher than the removal rates obtained in the semi-continuous reactor (TN 29.74 mg L$^{-1}$·d$^{-1}$, TP 0.96 mg L$^{-1}$·d$^{-1}$, TOC 29.32 mg L$^{-1}$·d$^{-1}$ and TIC 3.91 mg L$^{-1}$·d$^{-1}$). Otherwise, both PSBRs were able to produce biomass up to 0.09 g L$^{-1}$ d$^{-1}$, more than two times fold the biomass produced by semi-continuous reactor (0.04 g L$^{-1}$ d$^{-1}$), while obtaining a biomass settleability of 86-92%. Furthermore, this study demonstrated that microbial composition could be controlled by nutrients loads, since the three reactors were dominated by different species depending on the nutritional conditions. Concerning polymers accumulation, carbohydrates achieved similar values in the three reactors (11%), while <0.5 % of polyhydrohybutyrates (PHB) was produced. Low values in polymers production could be related to the lack of presence of microorganisms as cyanobacteria that are able to accumulate carbohydrates/PHB.

1. **Introduction**

Wastewater treatment with microalgae is regarded as an economical and environmentally friendly process with the additional advantage that the biomass produced can be reused and allows efficient nutrient recycling (Rawat et al., 2011; Honda, et al., 2012). In this process, microalgae work in association with aerobic heterotrophic bacteria (Abed et al., 2009; Borde et al., 2003). Indeed, photosynthetic microorganisms produce molecular oxygen that is used as electron acceptor by bacteria to degrade organic matter. In return, bacteria release carbon dioxide during the mineralization process and complete the photosynthetic cycle (Muñoz and Guieysse, 2006). This kind of wastewater treatment has been used for a range of purposes such as the removal of nutrients, the reduction of both chemical and biochemical oxygen demand and also for the removal of other compounds (i. e. heavy metals) (Abdel-Raouf et al., 2012; de Godos et al., 2009; Honda et al., 2012; Wang et al., 2010). On the other hand, wastewater is nowadays considered the only economically viable source of water and nutrients for the production of microalgae biomass that can then be used for valuable byproducts generation (Pittman et al., 2011; Uggetti et al., 2014).

In spite of the benefits, microalgae-based wastewater treatment technologies face operational limitations and challenges due to the high costs involving biomass separation from the treated wastewater (Renuka et al., 2013; Trivedi et al., 2015; Udom et al., 2013). This fact implies the use of biomass harvesting processes, whose technics commonly employed increase the production cost by about 20–30% of the total cost (Molina-Grima et al., 2003; Renuka et al., 2013; Yaakob et al., 2014). Recently, several studies have proposed to include a sedimentation period in the operational mode in order to increase spontaneous flocculation and the subsequent formation of big flocs (Valigore et al., 2012; Van Den Hende et al., 2016,

2014). This process can be carried out in a photo-sequencing batch reactor (PSBR), where hydraulic retention time (HRT) and solids retention time (SRT) are uncoupled as in activated sludge systems (Wang et al., 2015). In this way the cells are forced to form flocs that settle faster, while unsettling cells are removed from the supernatant (Valigore et al., 2012). Contrary to the conventional operations which do not promote extensive spontaneous flocculation (i.e. continuous, semi-continuous and batch), this approach can avoid additional intensive harvesting process. In addition, uncoupled HRT/SRT could influence nutritional dynamics and biomass composition. This can cause biochemical changes in microalgal biomass, affecting the accumulation of valuable polymers such as carbohydrates, lipids, and in the case of cyanobacteria polyhydroxybutyrates (PHBs) (Arcila and Buitrón, 2016; Arias et al., 2018). Those compounds have obtained increasing attention due to their potential use as biodiesel substrate, and in the case of PHBs as a bioplastics. The information of such promising alternative is still insufficient and all the aspects concerning nutrients dynamics in this kind of systems need to be addressed.

This study aims at comparing performances of sequencing batch and semi-continuous operations in terms of wastewater treatment, biomass composition and polymers accumulation. To this end, three photobioreactors (PBRs) were operated under SBR and semi-continuous mode as a tertiary treatment of digestate from anaerobic digester diluted with secondary wastewater from a high rate algal pond.

**2 Material and methods**

*2.1 Inoculum*

A mixed culture composed by green algae, cyanobacteria, bacteria, protozoa and small metazoa was used as inoculum. The thickened biomass (100 mL) was collected from a harvesting tank connected to a pilot closed-photobioreactor (30 L) already used as tertiary wastewater treatment (Arias et al., 2017).

*2.2 Experimental set-up*

Three lab scale photobioreactors consisting in closed polymethacrylate cylinders with an inner diameter of 11 cm with a total volume of 3 L and a working volume of 2.5 L each were used to perform the experiments. Reactors characteristics are detailed in a previous study (Arias et al., 2018). Experiments were carried out during 30 days.

The influent treated in the reactors consisted on uncentrifuged digestate diluted in secondary effluent from a high rate algal pond (HRAP) in a ratio of 1:50. Influent characteristics are shown in Table 1. The digestate was obtained from lab-scale anaerobic digesters (1.5 L) that produced biogas from microalgae biomass harvested from the HRAP. A detailed description of the system may be found in (Arias et al., 2018). The secondary effluent was obtained from a pilot system treating municipal wastewater which comprised a primary settler, a high rate algal pond (HRAP) and a secondary settler (Gutiérrez et al., 2016).

Table 1. Average (standard deviation) of the main water quality parameters of digestate, secondary effluent and the influent wastewater (constituted by digestate diluted in a ratio 1:50 with secondary effluent) (n=4).

| Parameter | Digestate | Secondary effluent | Influent wastewater |
|---|---|---|---|
| pH | - | - | 7.1 (0.8) |
| SST [$g \cdot L^{-1}$] | 21.85 (1.80) | -[a] | 0.44 (0.04) |
| SSV [$g \cdot L^{-1}$] | 17.90 (2.21) | -[a] | 0.36 (0.04) |
| TC [$mg \cdot L^{-1}$] | 20638.50 (1145.00) | 38.54 (6.00) | 413.23 (23.02) |
| TOC [$mg \cdot L^{-1}$] | 16993.5 (382.30) | 18.01 (3.20) | 340.23 (7.71) |
| TIC [$mg \cdot L^{-1}$] | 3645.00 (762.70) | 20.53 (2.8) | 73.31 (15.31) |
| TN [mg·L$^{-1}$] | 4685.41 (678.52) | 25.51 (5.98) | 83.35 (13.69) |
| TAN [mg·L$^{-1}$] | 1020.45 (233.99) | 0.045 (0.00) | 20.41 (4.68) |
| N-NO$_3^-$ [mg·L$^{-1}$] | <LOD | 8.99 (1.24) | 8.99 (1.24) |
| N-NO$_2^-$ [mg·L$^{-1}$] | <LOD | 1.22 (0.29) | 1.22 (0.29) |
| TIN [mg·L$^{-1}$] | 1020.45 (306.55) | 10.25 (3.45) | 30.62 (6.20) |
| TON [mg·L$^{-1}$] | 2644.51 (373.52) | 5 (1) | 52.99 (7.49) |
| TP [mg·L$^{-1}$] | 402 (115) | 3.22 (1.02) | 11.26 (1.63) |
| IP [mg·L$^{-1}$] | <LOD | 1.72 (0.13) | 1.72 (0.13) |
| TOP [mg·L$^{-1}$] | 402 (115) | 1.51 (0.60) | 9.54 (2.35) |

[a] TSS and VSS in the secondary effluent corresponded to values lower than 0.07 g L$^{-1}$.

All the reactors were continuously maintained in alternate light:dark phases of 12 h. Illumination during the light phase was supplied by two external halogen lamp (60W) placed at opposite sides of each reactor and providing 220 µmol m$^{-2}$ s$^{-1}$ of light. Reactors were continuously agitated (with the exception of settling periods) with a magnetic stirrer (Selecta, Spain) set at 250 rpm. Temperature was continuously measured by a probe inserted in the PBR (ABRA, Canada) and kept constant at 27 (±2) °C by means of a water jacket around the reactor. pH was continuously monitoring with a pH sensor (HI1001, HANNA, USA) and kept at 8.5 with a pH controller (HI 8711, HANNA, USA) by the automated addition of HCl 0.1 N or NaOH 0.1 N. Mixed liquor, supernatant, and feeding were performed by the automatic peristaltic pumps.

Two of the reactors were operated in a sequencing batch operation mode with a hydraulic retention time (HRT) of 2 days. One of those reactors (named PSBR$_{2-10}$) was operated with a solids retention time (SRT) of 10 days. This means that 0.25 L of mixed liquor were discharged at the end of the dark phase, successively the agitation was stopped and biomass was allowed to settle during 30 minutes. After this period, 1 L of the supernatant was withdrawn and then the total volume discharged (1.25 L) was replaced with the same volume of wastewater influent. The other sequencing batch reactor (named PSBR$_{2-5}$) was operated with a SRT of 5 days. Thus, 0.5 L of the mixed liquor were withdrawn at the end of the dark phase before a posterior settling time of 30 minutes. After the settling period, 0.75 L of the supernatant was withdrawn and then the total volume retired (1.25 L) was replaced with the same volume of wastewater influent. The operation of these PSBRs was compared with a semi-continuous reactor named $SC_{10\text{-}10}$ (control reactor). This last reactor was fed once a day and operated with a HRT and SRT of 10 days. This means that each day at the end of the dark phase, 0.2 L of the mixed liquor were withdrawn and subsequently this volume was replaced by 0.2 L of wastewater influent.

*2.3 Analytical methods*

*2.3.1 Nutrients concentrations*

Nutrients monitoring was carried out by analyzing samples taken from the reactors at the end of the dark phase after settling. All parameters were determined in triplicate and analyzed from the influent (mixed digestate and secondary effluent) and the supernatant of each reactor. Note that in the case of the reactor $SC_{10\text{-}10}$, the supernatant sample was taken from the mixed liquor withdrawn and submitted to a separation process. Samples from the influent were measured once per week, while in the samples of supernatant were analyzed three days per week.

Nitrogen was measured as total ammoniacal nitrogen (TAN), nitrite ($N\text{-}NO_2^-$), nitrate ($N\text{-}NO_3^-$), total nitrogen (TN) and total phosphorus (TP). TAN (sum of $N\text{-}NH_3$ and $N\text{-}NH_4^+$) was determined using the colorimetric method indicated in Solorzano (1969). $N\text{-}NO_2^-$ and $N\text{-}NO_3^-$ concentrations were analyzed using an ion chromatograph DIONEX ICS1000 (Thermo-scientific, USA), while TN was analyzed by using a C/N analyzer (21005, Analytikjena, Germany). Total inorganic nitrogen (TIN) was calculated as the sum of N-

$NO_2^-$, $N-NO_3^-$ and TAN. Total organic nitrogen (TON) (in dissolved and particulate form) was calculated as the difference between TN and TIN.

Phosphorus compounds analyzed were inorganic phosphorus (IP) measured as orthophosphate (dissolved reactive phosphorus) ($P-PO_4^{3-}$) and total phosphorus (TP). IP concentrations were analyzed using an ion chromatograph DIONEX ICS1000 (Thermo-scientific, USA) and total phosphorus (TP) was analyzed following the methodology described in Standard Methods (APHA-AWWA-WPCF, 2001). Total organic phosphorus (TOP) forms (dissolved and particulate) were calculated as the difference between TP and IP.

Total organic carbon (TOC), Total inorganic carbon (TIC), soluble organic carbon (OC) and soluble inorganic carbon (IC) were measured from raw and filtered samplesby using a C/N analyzer (21005, Analytikjena, Germany).

The volumetric load (Lv-X) of each nutrient (TOC, TIC, TAN, $NO_2^-$, $N-NO_3^-$, TIN, TON, TN, IP, TOP and TP) was calculated in [mg X $L^{-1}d^{-1}$] as follows:

$$Lv - X = \frac{Q * X}{V}$$

Where Q is the flow [$L^{-1}d^{-1}$], X is the nutrient influent concentration [mg X $L^{-1}$] and V [$L^{-1}$] is the volume of the reactor.

*2.3.2 Biomass concentration*

Total suspended solids (TSS) and volatile suspended solids (VSS) were measured in the mixed liquor at the end of the dark phase three days per week. In PSBR$_{2-10}$ and PSBR$_{2-5}$, two samples were taken; one from the mixed liquor right before to stop the agitation to evaluate the biomass production and one from the supernatant after the sedimentation to evaluate the settleability. Chlorophyll *a* was analyzed twice per week in the mixed liquor. Both analyzes procedures were performed by using the methodology described in the Standard Methods (APHA-AWWA-WPCF, 2001).

Biomass production of each reactor in [g VSS L$^{-1}$d$^{-1}$] was estimated following:

$$\textbf{Biomass production} = \frac{Q*VSS}{V}$$

where Q is the flow [L$^{-1}$d$^{-1}$], VSS is the biomass concentration in the reactor [g L$^{-1}$] and V [L$^{-1}$] is the volume of the reactor.

Settleability [%] was determinate according to the following formula:

$$\textbf{Settleability} = \textbf{100} * [\textbf{1} - \left(\frac{TSSs}{TSSm}\right)]$$

Where TSS$_m$ [mg L$^{-1}$] is the mixed liquor suspended solids concentration and TSS$_s$ [mg L$^{-1}$] is the supernatant suspended solids concentration.

Microalgae composition was monitored by a qualitative evaluation through microscope observations twice per week performed by an optic microscope (Motic, China) equipped with a camera (Fi2, Nikon, Japan) connected to a computer (software NIS-Element viewer®). Cyanobacteria and microalgae species were identified *in vivo* using conventional taxonomic books (Bourrelly, 1985; Palmer, 1962), as well as a database of Cyanobacteria genus (Komárek and Hauer, 2013).

*2.3.3 Polymers quantification*

Carbohydrates and polyhydroxybutyrates (PHB) content were measured twice per week in the biomass sampled from each reactor at the end the dark phase before the settling period. Then, 50 mL of mixed liquor were collected and centrifuged (4200 rpm,10 min), frozen at −80 °C overnight in an ultra-freezer (Arctiko, Denmark) and finally freeze-dried for 24 h in a lyophilizer (−110 °C, 0.049 hPa) (Scanvac, Denmark). PHB and carbohydrates extraction and quantification was performing the methodology described in Arias et al. (2018).

## 3. Results and discussion

*3.1 Nutrients dynamics and removal efficiency*

Due to the different HRT, nutrients volumetric load applied to $PSBR_{2-10}$ and $PSBR_{2-5}$ was five times fold higher than the load applied to $SC_{10-10}$ (Table 2). Furthermore, it is noticeable that the organic forms of nitrogen and phosphorus were the main sources of nutrients for the biomass. This fact is a direct consequence of the high TON and TOP contained in the digestate (Table 1) and thus, influence the TN and TP uptake and removals efficiencies.

Table 2. Nutrients volumetric load (Lv) in each reactor according to the hydraulic retention time (n=4).

| Parameter | $SC_{10-10}$[a] | $PSBR_{2-10}$[b] | $PSBR_{2-5}$[c] |
|---|---|---|---|
| Lv-TC [mg·L$^{-1}$·d$^{-1}$] | 41.35 (2.3) | 186.10 (9.36) | 186.10 (9.36) |
| Lv-TOC [mg·L$^{-1}$·d$^{-1}$] | 34.02 (0.77) | 153.11 (3.47) | 153.11 (3.47) |
| Lv-TIC [mg·L$^{-1}$·d$^{-1}$] | 7.33 (1.53) | 32.99 (6.89) | 32.99 (6.89) |
| Lv-TN [mg·L$^{-1}$·d$^{-1}$] | 8.65 (1.99) | 37.60 (8.95) | 37.60 (8.95) |
| Lv-TAN [mg·L$^{-1}$·d$^{-1}$] | 2.04 (0.47) | 9.18 (2.10) | 9.18 (2.10) |
| Lv-N-NO$_3^-$ [mg·L$^{-1}$·d$^{-1}$] | 0.90 (0.12) | 4.04 (0.55) | 4.04 (0.55) |
| Lv-N-NO$_2^-$ [mg·L$^{-1}$·d$^{-1}$] | 0.12 (0.03) | 0.55 (0.13) | 0.55 (0.13) |
| Lv-TIN [mg·L$^{-1}$·d$^{-1}$] | 3.06 (0.62) | 13.77 (2.79) | 13.77 (2.79) |
| Lv-TON [mg·L$^{-1}$·d$^{-1}$] | 5.29 (0.75) | 23.82 (3.37) | 23.82 (3.37) |
| Lv-TP [mg·L$^{-1}$·d$^{-1}$] | 1.13 (0.16) | 5.63 (0.82) | 5.63 (0.82) |
| Lv-IP [mg·L$^{-1}$·d$^{-1}$] | 0.17 (0.01) | 0.86 (0.07) | 0.86 (0.07) |
| Lv-TOP [mg·L$^{-1}$·d$^{-1}$] | 0.95 (0.24) | 4.77 (1.18) | 4.77 (1.18) |

As it can be observed in Fig. 1, TN in the effluent (without the biomass) showed similar concentrations in the three reactors. However, when comparing semi-continuous reactor with sequencing batch it is noticeable that the best performance in terms of nutrients assimilation and removal was reached by the sequencing batch operation (PSBR$_{2-10}$ and PSBR$_{2-5}$). Indeed, considering the higher load applied to the sequencing batch reactors (Fig. 1b and 1c), these showed a higher removal rates of TN (>29 mg L$^{-1}$ d$^{-1}$) than semi-continuous reactor (6.70 mg L$^{-1}$ d$^{-1}$) (Fig. 1a). It is important to remark that Lv-TN was constituted by 63% of TON and 37% of TIN (Table 1). Since it is impossible for microalgae to uptake organic nitrogen, TON should have been mineralized to TAN before to be consumed by microalgae (Pehlivanoglu and Sedlak, 2004). As observed in the three reactors, TON was almost totally mineralized, while TAN presented high variability along the time ranging from 0 to 13.45 mg L$^{-1}$ (Fig. 1). This suggests that high concentrations of TAN could be caused by the mineralization of TON. Regarding N-NO$_3^-$, it can be seen that the three reactors showed similar concentrations along the experiment (around 12 mg L$^{-1}$) (Table 3). In this case, similar concentrations in N-NO$_3^-$ are indicative of the higher assimilation of the reactors. While N-NO$_2^-$ also showed higher values in reactors than the influent, 3.84±3.33 mg L$^{-1}$ in SC$_{10-10}$, 6.08±4.52 in PSBR$_{2-10}$ and 6.63±4.28 mg L$^{-1}$ in PSBR$_{2-5}$. These high values of N-NO$_2^-$ observed in the three reactors suggest an inhibition in the nitrification process (Pollice et al., 2002).

In a wastewater treatment context, due to the similar TN concentrations in the three reactors (Table 3), similar removal percentages were obtained (63.3±0.6) (Table 4). Furthermore, high removals were observed in TAN (>80%) and TON (99%), while $N-NO_2^-$ and $N-NO_3^-$ were not removed in any reactor. In spite of such similarities in the general performance, the two PSBRs achieved more than 4 times higher removal rates in TN, TAN and TON (Table 4).

Table 3. Average (standard deviation) of the main nutrients concentrations of the supernatant of $SC_{10-10}$, $PSBR_{2-10}$ and $PSBR_{2-5}$ during the experiment (n=9-15).

| Parameter | $SC_{10-10}$[a] Average | $PSBR_{2-10}$[b] Average | $PSBR_{2-5}$[c] Average |
|---|---|---|---|
| IC [mg·L$^{-1}$] | 28.61 (23.69) | 39.60 (18.99) | 47.47 (20.77) |
| OC [mg·L$^{-1}$] | 47.41 (9.80) | 54.50 (23.46) | 49.84 (10.76) |
| TN [mg·L$^{-1}$] | 21.66 (7.12) | 23.59 (6.89) | 21.92 (4.96) |
| TAN [mg·L$^{-1}$] | 4.10 (5.08) | 3.71 (4.19) | 2.82 (3.25) |
| $N-NO_2^-$ [mg·L$^{-1}$] | 3.85 (3.33) | 6.08 (4.52) | 6.63 (4.28) |
| $N-NO_3^-$ [mg·L$^{-1}$] | 13.53 (4.78) | 12.33 (3.43) | 12.12 (4.48) |
| TIN [mg·L$^{-1}$] | 21.47 (7.20) | 22.12 (8.07) | 21.57 (5.62) |
| TON [mg·L$^{-1}$] | 0.19 (0.63) | 1.47 (2.93) | 0.035 (1.17) |
| TP [mg·L$^{-1}$] | 10.88 (2.89) | 14.63 (5.71) | 9.33 (6.69) |
| IP [mg·L$^{-1}$] | 1.37 (1.05) | 1.13 (1.41) | 2.90 (2.90) |
| TOP [mg·L$^{-1}$] | 6.89 (8.48) | 13.5 (4.30) | 6.43 (6.61) |

[a] Reactor operated with a coupled HRT and SRT of 10 d.
[b] Reactor operated with an uncoupled HRT of 2 days and SRT of 10 d.
[c] Reactor operated with an uncoupled HRT of 2 days and SRT of 5 d.

Table 4. Nutrients removal performances and removal rate of the effluent of the three reactors during the experiment (n=9-15).

| Parameter | SC$_{10-10}$[a] | | PSBR$_{2-10}$[b] | | PSBR$_{2-5}$[c] | |
| --- | --- | --- | --- | --- | --- | --- |
| | Removal percentage [%] | Removal rate [mg·L$^{-1}$·d$^{-1}$] | Removal percentage [%] | Removal rate [mg·L$^{-1}$·d$^{-1}$] | Removal percentage [%] | Removal rate [mg·L$^{-1}$·d$^{-1}$] |
| TOC | 86 | 29.32 | 84 | 128.78 | 85 | 130.81 |
| TIC | 53 | 3.91 | 40 | 13.13 | 35 | 11.63 |
| TN | 64 | 6.70 | 63 | 29.82 | 63 | 29.74 |
| TAN | 80 | 1.63 | 82 | 7.51 | 86 | 7.91 |
| N-NO$_3^-$ | - | - | - | - | - | - |
| N-NO$_2^-$ | - | - | - | - | - | - |
| TIN | 32 | 0.98 | 30 | 4.10 | 29 | 4.02 |
| TON | 99 | 5.29 | 99 | 23.58 | 99 | 23.58 |
| TP | 27 | 0.31 | - | - | 17 | 0.96 |
| IP | 20 | 0.03 | 34 | 0.29 | - | - |
| TOP | 29 | 0.27 | - | - | 33 | 1.56 |

[a]Reactor operated with a coupled HRT and SRT of 10 d.
[b]Reactor operated with an uncoupled HRT of 2 days and SRT of 10 d.
[c]Reactor operated with an uncoupled HRT of 2 days and SRT of 5 d.

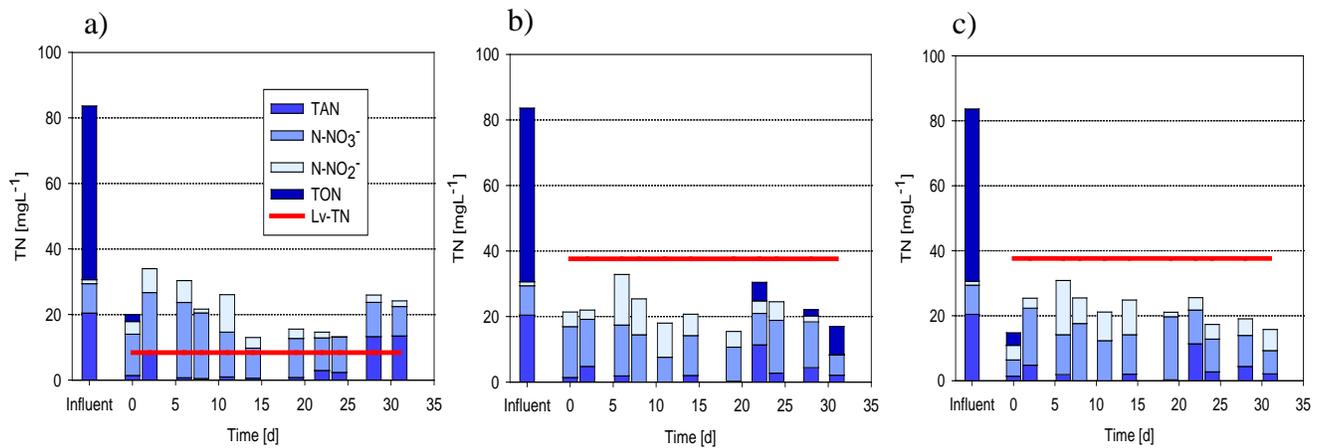

Figure 1. Average influent and effluent TN concentrations during the experiment in a) SC$_{10-10}$, b) PSBR$_{2-10}$ and c) PSBR$_{2-5}$. The average Lv-TN is presented in mg L$^{-1}$ d$^{-1}$.

On the other hand, TP in the effluent showed different patterns than those observed for TN. In general, the best performance was obtained in the semi-continuous reactor (SC$_{10-10}$) where the Lv-TP was very low (1.13±0.16 mg $L^{-1}$ $d^{-1}$) and was removed with a rate of 0.30 mg $L^{-1}$ $d^{-1}$. Otherwise, TP concentration in $PSBR_{2-10}$ showed an increasing pattern along the experimental time and values up to 15 mg $L^{-1}$ were reached in the last week of operation (Fig. 2b). In the case of $PSBR_{2-5}$, TP maintained concentrations higher than 10 mg $L^{-1}$ followed by a decrease to around 6 mg $L^{-1}$ in the two following weeks of operation (Fig. 2c).

These patterns in the three reactors were depending of the assimilation of TOP in all the reactors (Fig. 2). As for TON, microalgae are also unable to uptake organic phosphorus, then it is necessary that a mineralization process occurs to transform it to inorganic phosphorus species (Donald et al., 2017; Rodríguez and Fraga, 1999). High IP concentrations observed in several days indicate that TOP transformed to IP was not consumed. This fact can be clearly observed in $SC_{10-10}$ and in the last two weeks of $PSBR_{2-5}$ performance. Additionally, better assimilation of TOP was observed in $PSBR_{2-5}$ even though both PSBRs received the same Lv-TP. This fact can be related to the SRT of the reactors, since the best assimilation of TOP was performed by the reactor operating at 5 days. It is known that the mineralization process is microorganism dependent (Rodríguez and Fraga, 1999), this means that microalgae and bacteria growing with lower SRT were able to consume more P than the microorganism growing in a SRT of 10 days.

From the point of view of a wastewater treatment, the best TP removal efficiency percentages were achieved in $SC_{10-10}$ and $PSBR_{2-5}$ (Table 4). However, $PSBR_{2-5}$ showed a removal rate of TP of 1.56 mg $L^{-1}$ $d^{-1}$, which is six times higher the removal rate of $SC_{10-10}$. Due to the increased observed in TOP in $PSBR_{2-10,}$ no net removal was observed in this reactor.

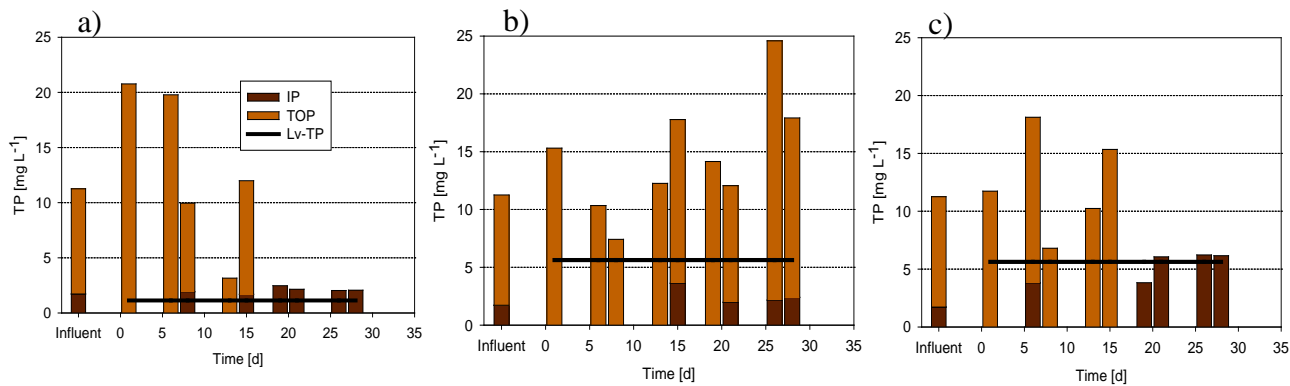

Figure 2. Average influent and effluent TP concentration during the experiment in a) $SC_{10\text{-}10}$, b) $PSBR_{2\text{-}10}$ and c) $PSBR_{2\text{-}5}$. The average Lv-TP is presented in mg $L^{-1}$ $d^{-1}$.

Regarding carbon forms uptake, although the three reactors averaged similar concentrations (Table 3), reactors showed differences in the assimilation in comparison with the Lv-TOC (Fig. 3). Due to the similar concentrations of TOC in the three reactors, removal efficiencies were similar (85±1%) with respect to the influent wastewater total content. However, removal rates in both PSBRs ($PSBR_{2\text{-}10}$ 128.78 mg $L^{-1}$ $d^{-1}$ and $PSBR_{2\text{-}5}$ 130.81 mg $L^{-1}$ $d^{-1}$) indicate a 4 times the rate removed with the semi-continuous reactor (29.32 mg $L^{-1}$ $d^{-1}$) (Table 4). In the case of TIC, although not eliminated, the effluent concentrations showed assimilation of this nutrient along the experiment (Fig. 3). In general, semi-continuous reactor showed the best TIC removal percentages (53%) (Table 4), notwithstanding, both PSBRs reached up to three times higher removal rates.

Throughout the results obtained, it is clear that the operation of $PSBR_{2\text{-}10}$ and $PSBR_{2\text{-}10}$ may be an alternative for the treatment of uncentrifuguous digestate diluted with secondary effluent within microalgal wastewater treatment systems. According to the removal rates, both PSBRs achieved the highest removals of TN, TOC and TIC, and TP (with the exception of $PSBR_{2\text{-}10}$). Nevertheless, due to the nitrification and the accumulation of TOP, the effluent is still out of the limits of the wastewater treatment standards (TN 15 mg L$^{-1}$ and TP 2 mg L$^{-1}$) (Directive 98/15/EC, 1998). Note that the limits presented in the normative are applied for urban wastewater treatment plants in communities between 10000-100000 p. e. According to the higher assimilation of TOP in SC$_{10-10}$, the increase in the HRT in the PSBRs could be a strategy to achieve a better assimilation of this compound and further research could be addressed to accomplish it. In the case of other nutrients assimilation, it was demonstrated that the PSBRs showed a better performance in relation to the load applied. Moreover such systems have the advantage that higher wastewater volumes can be treated per day.

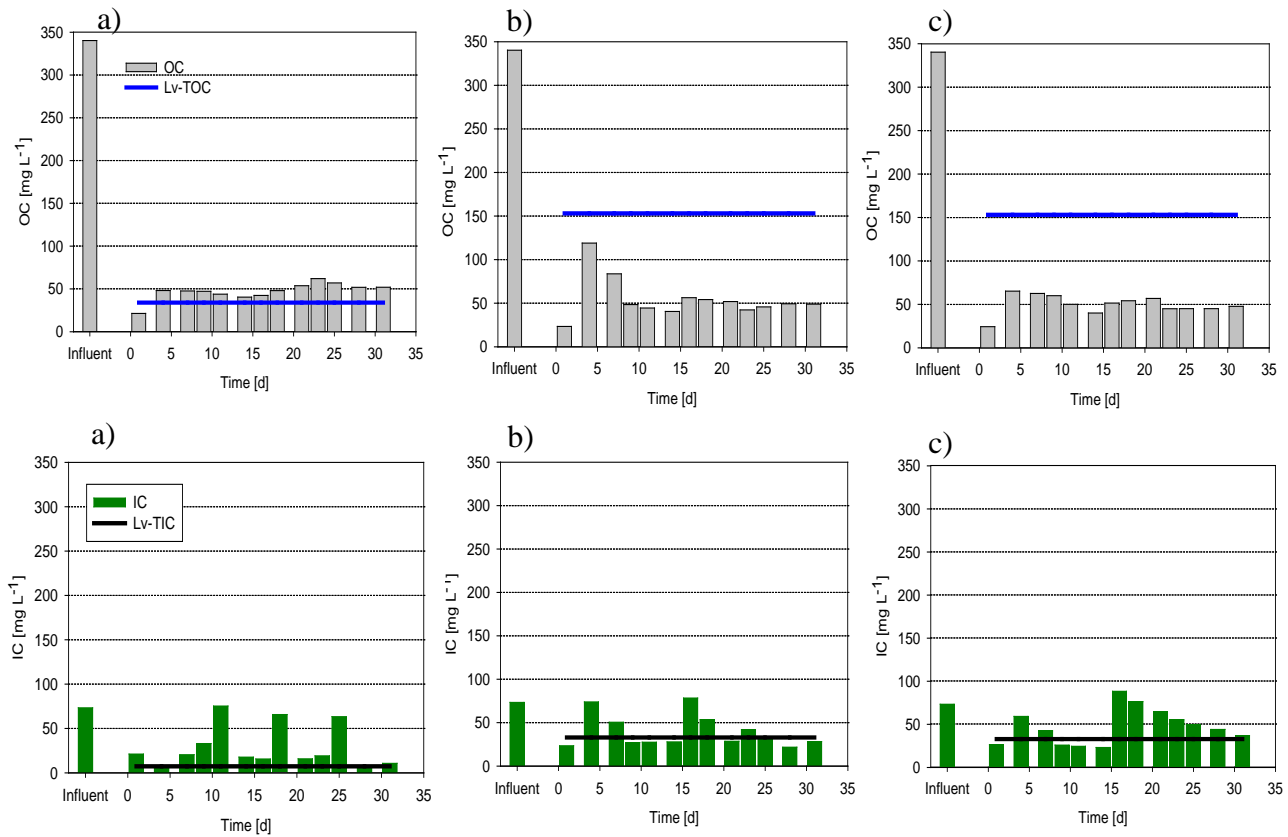

Figure 3. Average TOC and TIC influent and effluent OC and IC concentration during the experiment in a) SC$_{10-10}$, b) PSBR$_{2-10}$ and c) PSBR$_{2-5}$. The average Lv-TOC/TIC is presented as mg L$^{-1}$ d$^{-1}$.

*3.2 Biomass production*

In respect of biomass concentration, all the reactors showed an exponential increase in the first two weeks of operation. $SC_{10-10}$ increased from the initial concentration of 0.21±0.08 mg $L^{-1}$ to 0.451 mg $L^{-1}$ in day 15, and after this day it maintained a constant biomass concentration of approximately 0.420 mg $L^{-1}$. $PSBR_{2-10}$ maintained an increasing pattern until day 27 achieving the highest concentration of 0.910 mg $L^{-1}$. While $PSBR_{2-5}$ increased to 0.652 mg $L^{-1}$ in day 13 and subsequently decreased and maintained oscillating between 0.434 and 0.586 mg $L^{-1}$ along the experiment. With respect to the chlorophyll *a* content, $SC_{10-10}$ and $PSBR_{2-5}$ maintained a constant concentration along the experiment (0.597±0.091 and 0.829±0.279 mg $L^{-1}$, respectively) (Fig. 4), while $SPBR_{2-10}$ showed and increase from the initial concentration of 0.633 mg $L^{-1}$ to 2.82 mg $L^{-1}$ in the day 30.

In spite of the clear patterns registered in biomass concentration, the highest biomass production was achieved in $PSBR_{2-5}$ (Fig. 4), due to the highest volume withdrawn. Thus, the solids content production reached by this reactor was 0.135 mg $L^{-1}$ $d^{-1}$ in day 15, and as occurred in biomass concentration, it decreased in the following days maintaining a quite constant production of approximately 0.11 g VSS $L^{-1}·d^{-1}$. Despite the fact that in $PSBR_{2-10}$ a lower mixed liquor volume was extracted, in day 27 the biomass production achieved was similar to the one reached in $PSBR_{2-5}$. Otherwise, $SC_{10-10}$ only increased biomass production from 0.021 to 0.04 g $L^{-1}$ $d^{-1}$ in day 10, and subsequently maintained similar values.

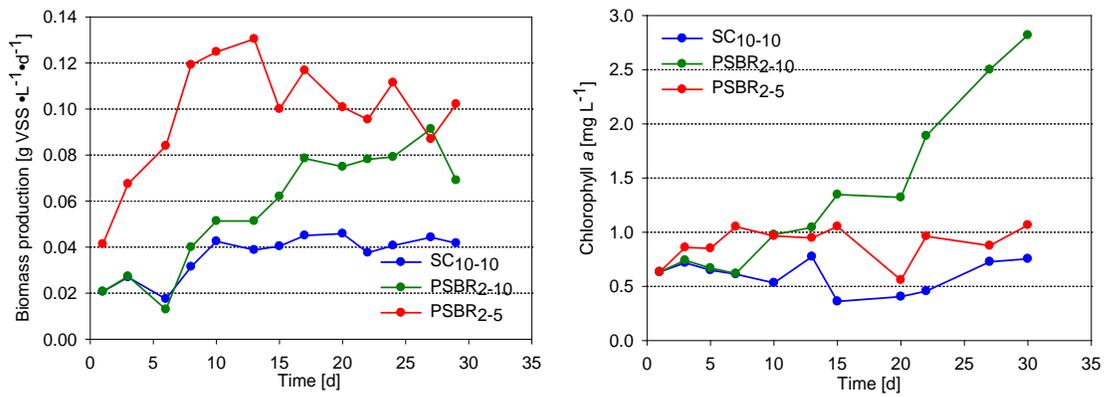

Figure 4. Time course of biomass production and chlorophyll *a* content.

According to the microscopic monitoring, $SC_{10-10}$ maintained similar microbial composition during the whole experiment (Fig. 5). The biomass was composed mostly by microalgal mixed flocs containing diatoms, unicellular cyanobacteria cf. *Aphanocapsa* sp., green algae species as *Chlorella* sp. and dispersed *Scenedesmus* sp., and rotifers protozoa. Bacterial colonies were also observed mostly in the last ten days of operation.

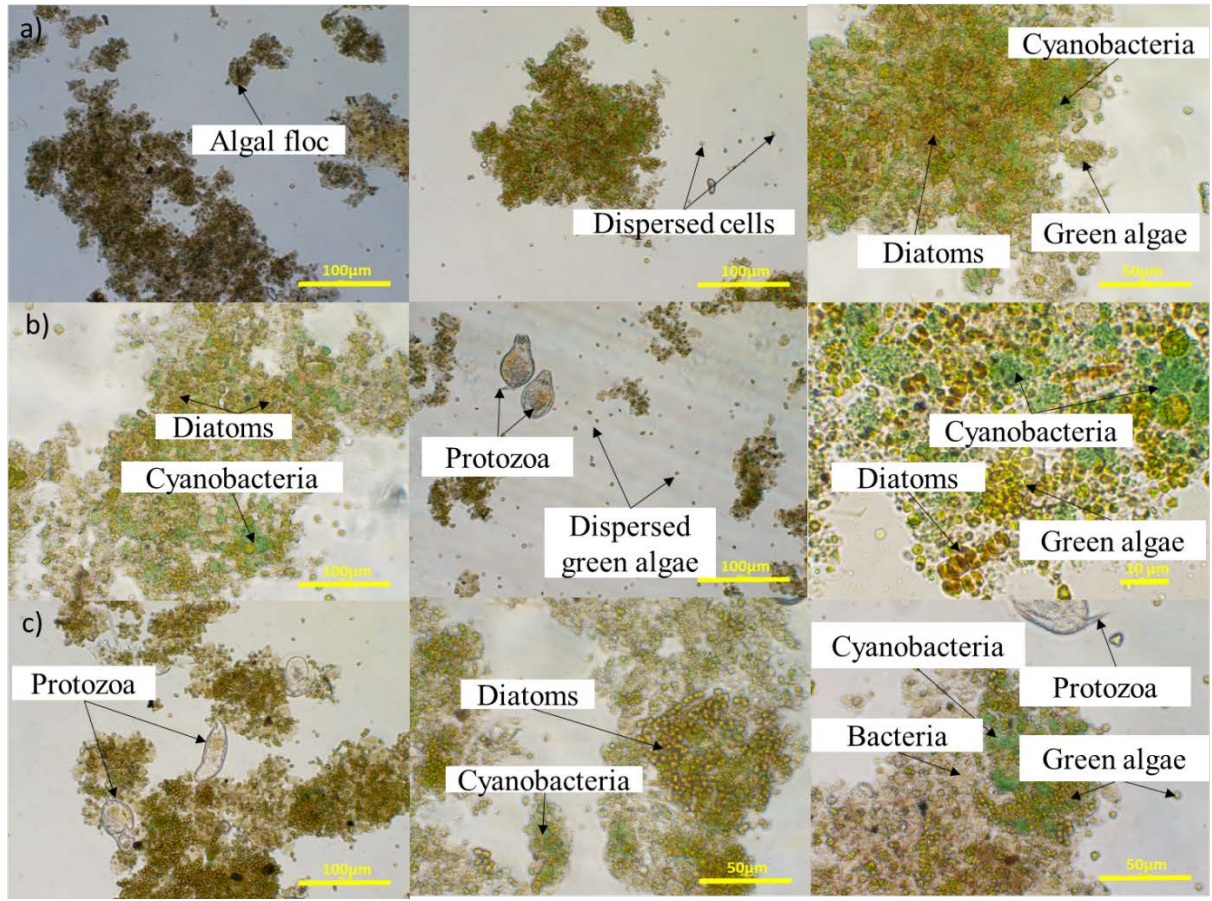

Figure 5. Microscopic images illustrating microbial composition in $SC_{10\text{-}10}$ during the periods; a) days 1-10, b) days 11-20 and c) days 21-30.

In $SPBR_{2\text{-}10}$ the first ten days showed a culture with the same composition observed in $SC_{10\text{-}10}$, with mixed flocs composed by green algae, some cyanobacteria and the presence of diatoms. However, microbial composition in posterior days showed an increasing presence of bacterial colonies (Fig. 6). Contrary to the reactor $SC_{10\text{-}10}$, $PSBR_{2\text{-}10}$ only increased green algae *Chlorella* sp. and dispersed cells of *Scenedesmus* sp. were not observed. Moreover, protozoa species as *Vorticella* sp. were frequently visualized.

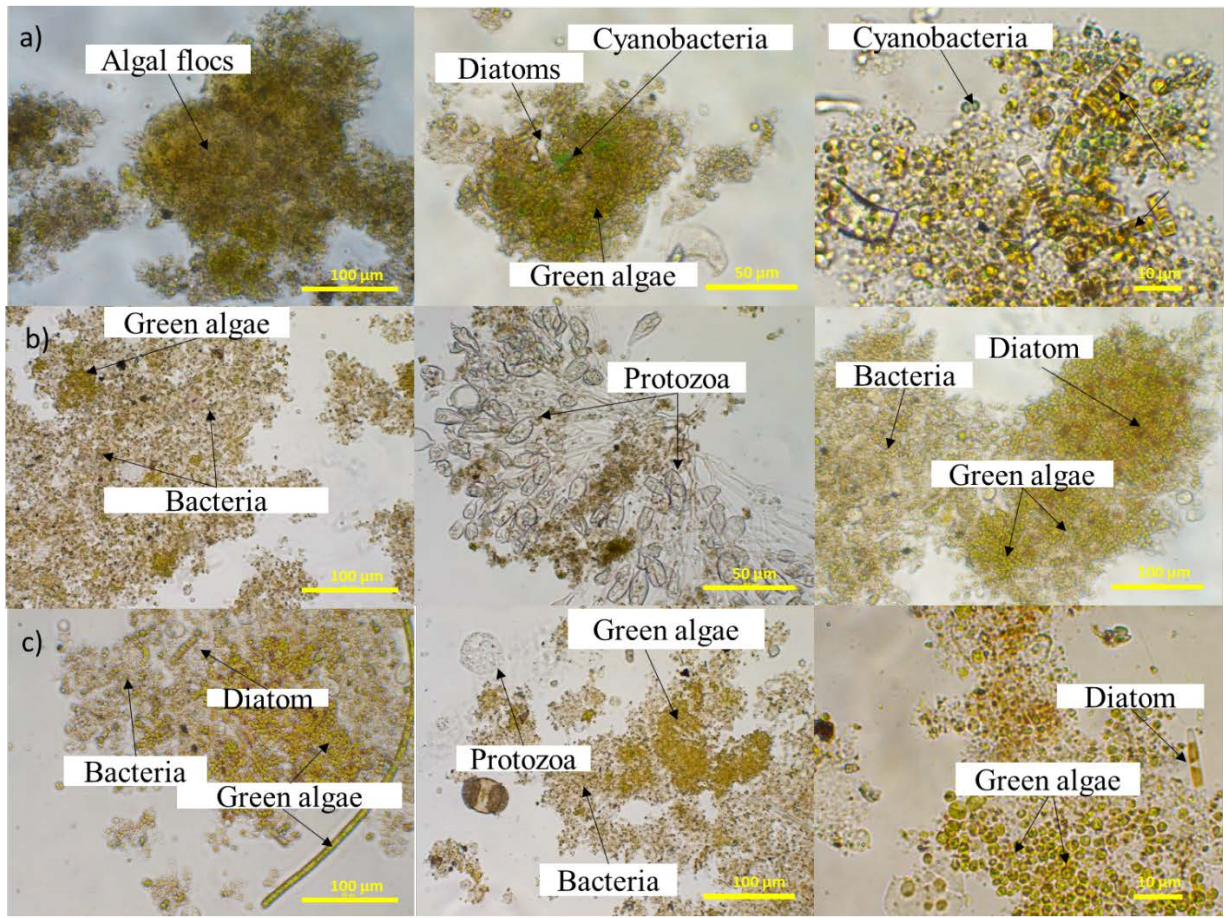

Figure 6. Microscopic images illustrating microbial composition in PSBR$_{2\text{-}10}$ during the periods; a) 1-10 days, b) 11-20 days and c) 21-30 days.

On the other hand, PSBR$_{2\text{-}5}$ showed a different microbial evolution in comparison to the other reactors. As observed in Fig. 7, algal flocs were rarely observed, instead, bacterial flocs were observed from the first days of operation onwards. In this reactor green algae present in the culture belongs to species of *Chlorella* sp. and *Stigeoclonium* sp.. Other species of protozoa, cyanobacteria and diatoms were rarely observed in the culture.

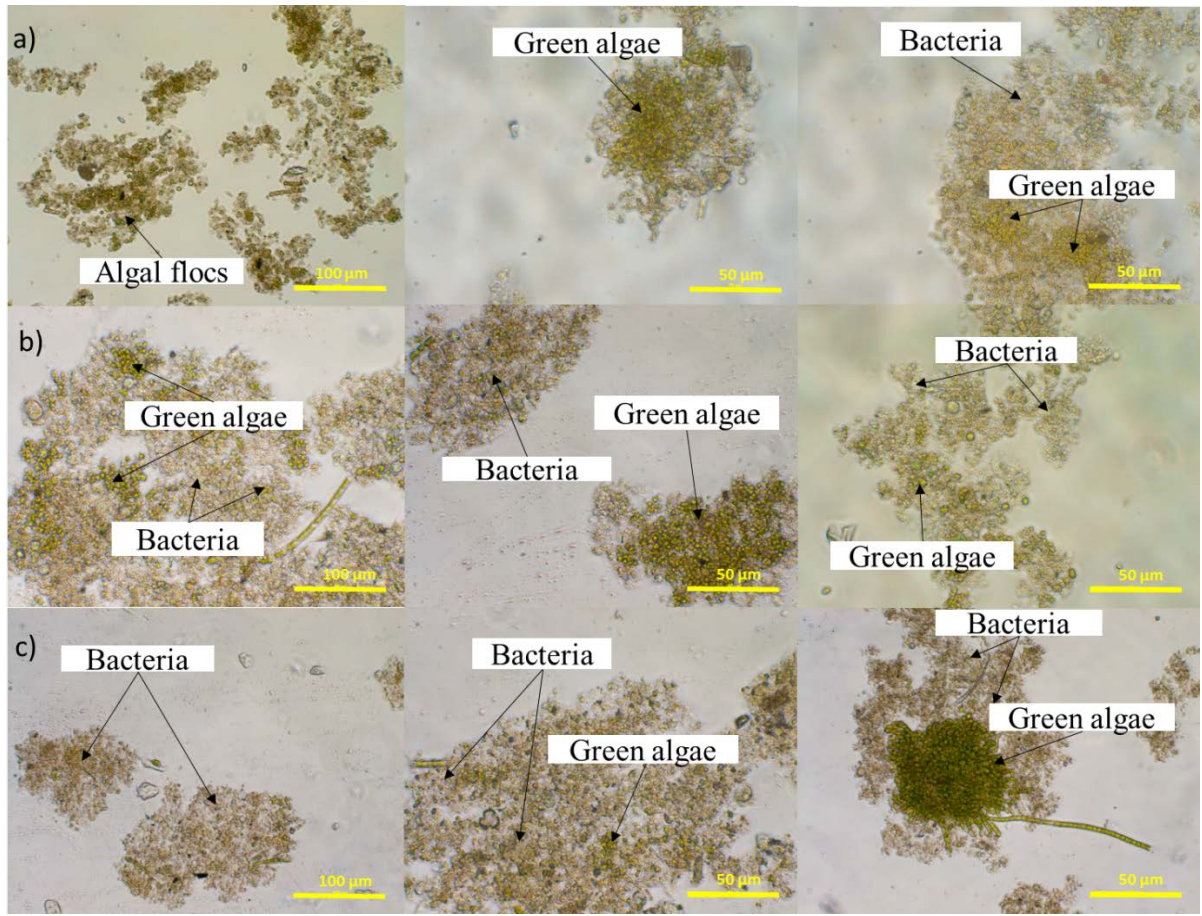

Figure 7. Microscopic images illustrating microbial composition in PSBR$_{2-5}$ during the periods; a) days 1-10, b) days 11-20 and c) days 21-30.

In addition to the lack of dispersed cells observed by microscopy in PSBR$_{2-10}$ and PSBR$_{2-5}$, SSV in the supernatant was maintaining 0.075±0.021 mg L$^{-1}$ and 0.072±0.003 mg L$^{-1}$, respectively, from the first days of operation and along the experimental time. Such values imply a settleability of 86 to 92%. When comparing the biomass composition of the three reactors, it is clear that the strategy of operating in PSBR$_{2-10}$ and PSBR$_{2-5}$ with uncoupled SRT and HRT improved capacity of the microorganisms to form flocs and perform a fast settling process, which is a good result from the harvesting point of view.

Considering this system for biomass production, this study demonstrated that microbial composition could be controlled by nutrients loads and at the same time the presence of certain microorganism are influenced by the SRT. In the case of protozoa and diatoms, this study showed that these microorganisms can survive in a wide range of loads since their presence was observed in either low loads ($SC_{10-10}$) or high loads ($PSBR_{2-10}$). However, their presence was conditioned to a long SRT of 10 days. Protozoa and diatoms are usually observed in this type of systems with long SRT (Shariati et al., 2011). Otherwise, the fact that cyanobacteria presence could occur in $SC_{10-10}$ (low loads) but not in $PSBR_{2-10}$ (high loads), even if they had the same SRT, showed that nutritional conditions highly affects cyanobacteria presence. However, it would be important to improve this specie competition capacity in microalgae-based wastewater treatments since they are potential PHB and carbohydrates producers.

Another important fact to take into account is that bacteria presence increased more in $PSBR_{2-10}$ and $PSBR_{2-5}$ than $SC_{10-10}$. This suggests that the introduction of high loads of nutrients, specially TOC in the PSBRs, promoted the growth of heterotrophic bacteria. While another important fact to be considered is that the influent used in this experiment (secondary effluent and digestate) contained high TOC:TIC ratio (4.64). Previous study of Van Den Hende et al., (2014) showed that TOC:TIC ratios higher than 2.39 improve heterotrophic bacteria domination in PSBRs operated at 2 days of HRT. However, the fact that the semi-continuous reactor with the same influent but with less nutrients load showed a dominance of microalgae suggests that load applied to the reactor also played an important role in the microbial community composition.

It is important to highlight that although microalgae based wastewater treatments similar to those described in this study, have been used successfully used for the treatment of digestates from different sources, the most of the studies until now have employed batch or semi-continuous operation (Cañizares-Villanueva et al., 1994; Pouliot et al., 1989; Ruiz-Marin et al., 2010; Sepúlveda et al., 2015; Uggetti et al., 2014; Viruela et al., 2016), which implies a limitation in nutrients removal rates, biomass production and the possibility to produce an easy settling culture. The strategy of sequencing batch operation of photobioreactors for digestate removal is still limited to only few studies. In the study of Van Den Hende et al., (2014), a 4 L PSBR operated at an HRT of 2 d to treat manure digestate was utilized. Within their achievements, removal rates of TN and TP of 4.5 mg $L^{-1}$ $d^{-1}$ and 0.11 mg $L^{-1}$ $d^{-1}$, respectively, were obtained, while producing 0.068 g $L^{-1}$ $d^{-1}$ of biomass. Remarkably, the results of $PSBR_{2-5}$ of this study reached higher removals rates, since 29.82 and 1.05 mg $L^{-1}$ $d^{-1}$ of TN and TP were removed, respectively, and at the same time a higher biomass production was achieved (0.11 g $L^{-1}$ $d^{-1}$). On the other hand, the removal rate of TN of this study was lower than the study of Wang et al., (2015), who used a 8 L PSBR operated at an HRT of 4 d for the removal of diluted digestate, removed 71 mg $L^{-1}$ $d^{-1}$ of TN employing nitrification and denitrification strategies in the PSBR, and at the same time produce 0.15 g $L^{-1}$ $d^{-1}$ of biomass.

*3.3 Polymers accumulation*

As mentioned in Section 3.1, in general none of the reactors presented nutrients limitation along the experiment. Due to this condition of high nutrients availability, low polymers accumulation occurred in the cultures. With respect to carbohydrates content, they only achieved low similar and constant content. Hence, $SC_{10-10}$ reached 11.18±1.76 % $VS^{-1}$, while

PSBR$_{2-10}$ and PSBR$_{2-5}$ achieved 11.47±2.78 and 9.90±2.60 % VS$^{-1}$. Despite the fact that the three reactors showed similar percentages, different concentrations were achieved considering the biomass concentrations: the highest concentration (128.60±13.69 mgL$^{-1}$) was achieved in PSBR$_{2-10}$, while PSBR$_{2-10}$ and PSBR$_{2-5}$ maintained a constant concentration of 53.11±10.04 mg L$^{-1}$ (Fig. 8). On the other hand, low PHB accumulation was observed in all the reactors during the experimental time (<0.5% PHB VS$^{-1}$).

Results of carbohydrates reached in this study showed lower content that the ones obtained by Arcila and Buitrón (2016) in a HRAP operated at hydraulic and solids retention times of 2, 6 and 10 d (12, 16 and 22%, respectively). It is important to remark that this study and the study of Arcila and Buitrón (2016) were conducted in absence of nutrients limitation, which is an important factor limiting the accumulation of carbohydrates (De Philippis et al., 1992; Markou et al., 2013).

The fact that PHB was not accumulated was caused by the lack of cyanobacteria in the cultures. As already explained in biomass evolution Section 3.2, reactors were mostly composed by green algae, which are not accumulating PHB. Thus, the low values of this polymer was expected since its accumulation is conditioned by a culture composed by PHB accumulating microorganisms (i. e. bacteria and cyanobacteria). Besides the fact that this polymer is accumulated during starving conditions of nitrogen or phosphorus (Arias et al., 2018; Samantaray et al., 2011). These two facts could likely influence the poor accumulation of this polymer in this study.

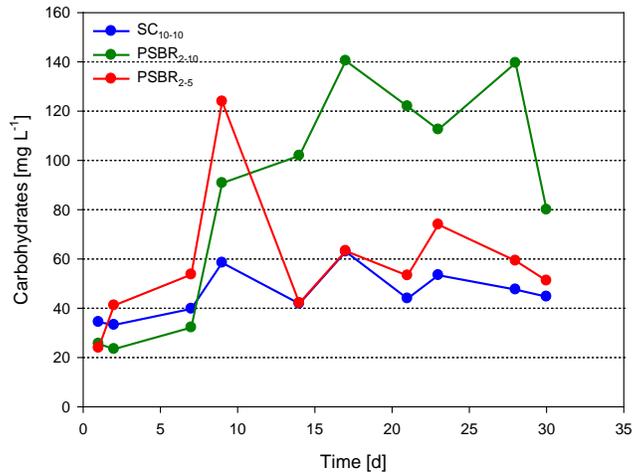

Figure 8. Time course of carbohydrates concentration.

## 4. Conclusions

In this study, nutrients removal and biomass growth were analyzed in photosynthetic sequencing batch reactors (PSBR) treating digestate diluted with secondary effluent. Two PSBR were operated at hydraulic retention time (HRT) of 2 days and solids retention time (SRT) of 10 and 5 days, and results were compared with semi-continuous (SC) reactor operating at HRT and SRT of 10 days. PSBR showed removals rates of 30 mg $L^{-1}$ $d^{-1}$ of total nitrogen and up to 1 mg $L^{-1}$ $d^{-1}$ of total phosphorus. Concerning inorganic carbon and organic carbon uptake, PSBRs achieved removals rates of 128-130 mg TOC $L^{-1}$ $d^{-1}$ and 12-13 mg TIC $L^{-1}$ $d^{-1}$. Those results were in general 1-5 times higher than the removal rates obtained in the semi-continuous reactor. Otherwise, PSBRs were able to produce biomass up to 0.09 g $L^{-1}$ $d^{-1}$, more than two times fold the biomass produced by SC, while obtaining a biomass settleability of 86-92%. Furthermore, this study demonstrated that microbial composition could be controlled by nutrients loads, since the three reactors were dominated by different species depending on the nutrients concentrations. Concerning polymers accumulation, carbohydrates achieved similar values in the three reactors by 11%, while <0.5 % of polyhydrohybutyrates (PHB) was produced. Low values in polymers production could be related to the lack of cyanobacteria which are the microorganism accumulating carbohydrates/PHB. Future studies should be also directed to determine nutrients strategies to select appropriated microorganisms and at the same time enhance polymers accumulation.

**Acknowledgments**

The authors would like to thank the European Commission [INCOVER, GA 689242] and the Spanish Ministry of Science and Innovation [project FOTOBIOGAS CTQ2014-57293-C3-3-R] for their financial support. Dulce Arias kindly acknowledge her PhD scholarship funded by the National Council for Science and Technology (CONACYT) [328365]. M.J. García and E. Uggetti would like to thank the Spanish Ministry of Industry and Economy for their research grants [FJCI-2014-22767 and IJCI-2014-21594, respectively]. Authors appreciate Estel Rueda and Laura Torres for their contribution during the experiment deployment.